\documentclass[prl,letterpaper,twocolumn,aps,epsf]{revtex4-1}
\usepackage{graphicx}
\usepackage{epsfig}
\def\comment#1{}

\newcommand{\beg}{\begin{eqnarray}}
\newcommand{\eee}{\end{eqnarray}}

\def\cm#1{}

\newcommand{\chiv}{\mbox{\boldmath{$\chi$}}}

\newcommand{\be}{\begin{equation}}
\newcommand{\ee}{\end{equation}}
\newcommand{\ba}{\begin{eqnarray}}
\newcommand{\ea}{\end{eqnarray}}
\newcommand{\beq}{\begin{equation}}
\newcommand{\eeq}{\end{equation}}
\newcommand{\bea}{\begin{eqnarray}}
\newcommand{\eea}{\end{eqnarray}}
\newcommand{\bastar}{\begin{eqnarray*}}
\newcommand{\eastar}{\end{eqnarray*}}

\newcommand{\cd}{\partial}
\newcommand{\hess}{{\cal H}}

\renewcommand{\tt}{}

\begin{document}
\title{Type-1.5 Superconducting State from an Intrinsic Proximity Effect in Two-Band Superconductors}

\author{
Egor Babaev${}^{1,2}$, Johan Carlstr\"om${}^1$ and  Martin Speight${}^3$}
\address{
${}^1$Department of Theoretical Physics, The Royal Institute of Technology, Stockholm, SE-10691 Sweden\\
${}^2$ Department of Physics, University of Massachusetts Amherst, MA 01003 USA \\
${}^3$ School of Mathematics, University of Leeds, Leeds LS2 9JT, UK
}

\begin{abstract}

We show that in multiband superconductors even small  interband proximity effect
 can lead to a qualitative change in the  interaction potential between superconducting vortices  by producing  long-range intervortex attraction. 
This type of vortex interaction results in unusual response to low magnetic fields leading to phase separation into domains of 
a two-component Meissner states and vortex droplets.

\end{abstract}
\maketitle

The textbook classification of superconductors, divides them into two classes,  according 
to their behavior in an external field. Type-I superconductors expel low magnetic fields,
while elevated fields produce macroscopic normal domains in the interior of the superconductor.
Type-II superconductors possess  stable vortex excitations
which can form a vortex lattice as the energetically preferred state in an applied magnetic field.
This picture of type-II superconductivity, as well
as the essence of the more complex physics of fluctuating vortex matter, 
relies on the fact that the interaction between co-directed vortices is purely repulsive.
In \cite{bs1} it was demonstrated that in $U(1)\times U(1)$  superconductors
with two independent components, in a wide parameter range,  there are vortex solutions 
which are on the one hand thermodynamically stable, and on the other hand, possess a non-monotonic interaction potential,
repulsive at short distances but attractive at larger distances.
Long range vortex attraction in the models \cite{bs1} originates from the
circumstance  that the coherence length of one of the components
is the largest length scale of the problem,
and the core of one of the components
extends to the region where current and magnetic field (which are
responsible for repulsive intervortex interactions)
are exponentially suppressed. 
Indeed such a
vortex interaction, along with their demonstrated thermodynamic stability, should cause the system to respond to external fields
in an entirely different way from the vortex states of traditional type-II superconductors. Namely, the 
attraction between vortices 
should, at low fields, produce the ``semi-Meissner
state" \cite{bs1}) featuring (i) formation of voids of vortex-less states, 
where 
there are two well developed superconducting components and (ii) {  vortex clusters
where the second component is  suppressed by  overlapping of vortex outer cores}.
This kind of external field-induced ``phase separation" 
which, from the point of view of the second  component, resembles a mixed state of type-I superconductors,
can be interpreted as the system showing aspects of type-I and type-II magnetic response simultaneously.
The term type-1.5 superconductivity was 
coined for this kind of behavior.
{  Note that this magnetic response 
originates from the existence of three fundamental
length scales in the problem (in contrast to 
the ratio $\kappa$ of
two fundamental length scales which parametrizes single-component Ginzburg-Landau theory),
and thus it is entirely different from  
the inhomogeneous vortex states in
single-component superconductors 
where inhomogeneity can be
induced by defects in a type-II superconductor or by tiny
attraction caused by various nonuniversal microscopic effects
beyond the Ginzburg-Landau theory which might be
pronounced in single-component superconductors with $\kappa$ is extremely close to $1/\sqrt{2}$ \cite{brand}.}

Recently there has been {strong and growing} interest in multi-band materials
where intercomponent interaction can be  substantial.
Examples are $MgB_2$  \cite{mgb,gurevich}
and possibly new iron-based superconductors \cite{iron}. 
The  two-band superconductor $\rm MgB_2$  \cite{mgb,gurevich} was regarded in 
 early theoretical and experimental works 
as a  standard type-II superconductor. This  was disputed in the recent
 works by Moshchalkov et al  \cite{m1,m2}, which reported
 highly inhomogeneous vortex states formation in clean samples in low magnetic fields
with vortex clusters (with a preferred intervortex separation scale) 
and vortex-less Meissner domains strikingly similar to the  picture of 
the semi-Meissner state \cite{bs1}.
 In connection with the experiments \cite{m1,m2} 
 and recent suggestions that iron pnictides may also be multi-component superconductors,
 the question arises under what conditions
type-1.5 superconductivity is possible (even in principle) in general multi-band systems
with a substantial interband coupling.

{  In this Letter we show  that type-1.5 behaviour can arise via a new mechanism in a two-band
system with a direct coupling between the bands.
The situation which we consider is, in a way,
{ antipodal}   to that considered in \cite{bs1}: namely where only one band is truly superconducting 
 while superconductivity in the other band
is induced by the interband  proximity effect. 
}We address the 
  properties of such a regime by 
studying the following free energy density
{ (in units where $\hbar=c=m=1$ and $e$ is the Cooper pair charge).}

\begin{eqnarray}
\label{ind_energy}
\mathcal{F}&=&  
\frac{1}{2}\sum_{i=1.2}|(\nabla+ ie{\bf A}) \psi_i  |^2+ 
\frac{1}{2}(\nabla \times {\bf A})^2   +\frac{1}{2}\Big(|\psi_1|^2-1\Big)^2 \nonumber \\
&+&\alpha|\psi_2|^2 +\frac{1}{2}\beta|\psi_2|^4 -\eta|\psi_1|| \psi_2|\cos(\theta_2-\theta_1) 
\end{eqnarray}
%
%
{ Here $\psi_{1,2}$ represent
 the superconducting  components associated with two bands.}
The radical difference
with previous studies \cite{bs1} is that in (\ref{ind_energy})
the effective potential for $\psi_2$ has only positive terms $ \alpha,\beta >0$, i.e.\ this band is {\it above} its critical temperature.
It has a nonzero density of Cooper pairs only because of the interband  tunneling represented by the term $-\eta|\psi_1|| \psi_2|\cos(\theta_2-\theta_1)$ (since the Josephson term favours locked phases we have $\theta_1=\theta_2\equiv \theta$). 
 The results can be  generalized to including other mixed gradient and density terms in (\ref{ind_energy}). 
 In what follows we will denote the ground state values of $|\psi_1|$ and $|\psi_2|$ by $u_1$ and $u_2$. 
Note that 
in this model, in general, no explicit expressions for $u_1$ and
$u_2$ in terms of $\alpha,\beta,\eta$ exist, but one can compute power series
expansions for them in $\eta$,
\beq
u_1=1+\frac{\eta^2}{8\alpha}+O(\eta^4),\qquad
u_2=\frac{\eta}{2\alpha}+O(\eta^3).
\eeq
Vortex solutions of the model take the form $\psi_a=\sigma_a(r)e^{i\theta}$,
${\bf A}=r^{-1}a(r)(-\sin\theta,\cos\theta)$, where $\sigma_a(\infty)=u_a$ and $a(\infty)=-e^{-1}$.
To understand the long-range behaviour of a vortex, we choose
gauge so that $\psi_1,\psi_2$ are real, set $\psi_i=u_i+\chi_i$, and linearize
the model about $\chiv=(\chi_1,\chi_2)^T=(0,0)^T$, ${\bf A}=0$. 
The result is a {\em coupled} Klein-Gordon
system with energy density
\beq
E=\frac12\left\{|\nabla\chiv|^2+\chiv^T\hess\chiv+|\nabla\times {\bf A}|^2+e^2(u_1^2+u_2^2)|{\bf A}|^2\right\},
\eeq
where $\hess$ is the Hessian of $V$ about the 
ground state $|\psi_i|=u_i$,
that is, $\hess_{ij}=
\cd^2V/\cd|\psi_i|\cd|\psi_j|$. Clearly,
\beq
\hess=\left(\begin{array}{cc}6u_1^2-2&-\eta\\
-\eta & 6\beta u_2^2+2\alpha\end{array}\right).
\eeq
The eigenvalues of $\hess$ are the squared masses of the normal modes
about the ground state.  If  $\eta=0$, then $\chi_1$ and $\chi_2$ decouple and have
masses $2$ and $\sqrt{2\alpha}$,  the first one being in this limit the inverse coherence
length of the first condensate, as expected.  If $\eta>0$,
both condensates have nonzero ground state values $u_1$ and $u_2$ which 
are not known explicitly, but importantly the normal modes are not $\chi_1$, $\chi_2$,
but rather an orthogonal pair $(\chi_1\cos\omega+\chi_2\sin\omega$,
$-\chi_1\sin\omega+\chi_2\cos\omega)$
where $(\cos\omega,\sin\omega)^T$ and $(-\sin\omega,\cos\omega)^T$ are 
the eigenvectors of $\hess$. Physically this means that the recovery of densities
in both bands from the core singularity has a  {  strong mutual dependence}.
 The London penetration length is given by the inverse mass of ${\bf A}$:  $\mu_A=e\sqrt{u_1^2+u_2^2}$.
So the linear theory predicts, at large $r$, the asymptotic formulae
\bea
|\psi_1|&\sim&u_1-q_1\cos\omega K_0(\mu_1 r)+q_2\sin\omega K_0(\mu_2 r)\nonumber\\
|\psi_2|&\sim&u_2-q_1\sin\omega K_0(\mu_1 r)-q_2\cos\omega K_0(\mu_2 r)\nonumber\\
|{\bf A}|&\sim&r^{-1}(e^{-1}+q_AK_1(\mu_A r))
\eea
where $K_m$ denotes the $m$-th modified Bessel function of the second kind, and $q_1,q_2,q_A$ are some unknown real constants
depending on $\alpha,\beta,\eta,e$. Recall that $K_m(r)\sim (\pi/ 2r)^\frac12e^{-r}$ for all $m$.
 { This means that, in spite of the presence
of two superfluid densities, we  {\it cannot} talk about two distinct coherence lengths (in the GL sense)
pertaining to these condensates: the leading term in both $|\psi_1|-u_1$ and $|\psi_2|-u_2$ decays
exponentially with the {\it same} length scale, $\xi=\max\{\mu_1^{-1},\mu_2^{-1}\}$}. 
{  At the same time the system retains three fundamenatal length scales, 
which in this case are the magnetic field penetration length and two inverse masses 
$\mu_1^{-1},\mu_2^{-1}$ of the modes
associated with density variation in the coupled bands.}
Applying the methods of \cite{spe}, one finds that the asymptotic interaction potential for two well-separated vortices
is
\beq\label{V}
V\sim 2\pi[q_A^2K_0(\mu_A r)-q_1^2K_0(\mu_1 r)-q_2^2 K_0(\mu_2 r)].
\eeq
The first term represents repulsion due to current-current and magnetic field interactions, while the last two
terms represent attractive forces 
associated with nontrivial density modulation, mediated  in this case by the normal modes,  described by scalar fields of mass $\mu_1$ and $\mu_2$.
{\tt Hence, the linearized theory predicts that vortices should attract at very large separations if 
$\min\{\mu_1,\mu_2\}<\mu_A $ (and
repel if $\min\{\mu_1,\mu_2\}>\mu_A $)}. It should be emphasized  that the above analysis concerns the leading asymptotics of the vortex
fields, not the core structure of the vortex directly. { As we discuss below,  in the presence 
of interband Josephson tunneling the detailed core structure is principally important
for the form of the  vortex interaction potential  (in contrast to usual 
single-component superconductors). }
This  core structure cannot be derived in a simple manner from eq.\ (\ref{ind_energy}). Rather, it
must be deduced from numerical solutions of the full, nonlinear field equations which are presented
in the second half of the paper.

 However for small $\eta$ we can demonstrate analytically that
vortices in the model (\ref{ind_energy})
can attract one another at long range
but  repel at  short range
by finding the masses  $\mu_A,\mu_1,\mu_2$.
That is, we can find expansions for
these masses and the ``mixing angle" $\omega$, valid for 
small $\eta$,
\bea
\mu_A&=&e+O(\eta^2);\ \mu_1=2+O(\eta^2); \  \mu_2=\sqrt{2\alpha}+O(\eta^2) \nonumber \\
 \omega &=&\frac{\eta}{|2\alpha-4|}+O(\eta^2) .
\eea
There is a range of parameters where $u_i,\mu_i$ and
$\omega$ can be computed explicitly, namely $\beta=0$. Physically, this
limit is sensible if $|\psi_2|$ remains everywhere small, which it does for
small $\eta$, since $u_2=O(\eta)$. One finds that
\bea
u_1&=&\sqrt{1+\frac{\eta^2}{4\alpha}};\ u_2=\frac{\eta}{2\alpha}u_1; \ \omega=\tan^{-1}\left|\frac{\eta}{2\alpha-\mu_1^2}\right|;  \nonumber\\
\mu_{1,2}^2&=&\frac{p\pm\sqrt{p^2-128\alpha^3-32\eta^2\alpha^2}}{4\alpha}
\eea
where $p=8\alpha+4\alpha^2+3\eta^2$.
To understand the  intervortex forces at short range we note that
for small $\eta$, $\psi_2$ remains close to zero
throughout the core of $\psi_1$ and in the most of the flux-carrying area, so one expects $\psi_2$ to
contribute negligibly to the interaction energy at short range.
In this limit one can approximate  the vortex solution 
at this scale by setting $\psi_2=0$ in $F$ yielding a 
one component GL model with GL
parameter $\kappa_{GL}=\sqrt{2}e^{-1}$, leading one to predict short range
vortex {\em repulsion} for $0<e<2$ for the effective potential given in eq. (\ref{ind_energy}). 
So, linear and qualitative analysis
suggests that the model (\ref{ind_energy}) 
 does possess type-1.5 superconductivity at least
whenever $0<e<2$ and
the condition for long range attraction holds, namely 
$\min\{\mu_1^2,\mu_2^2\}<e^2(u_1^2+u_2^2)$, where $\mu_i^2$
are the eigenvalues of $\hess$.
In order to test this prediction,
and to study regimes where analytic estimates cannot be made,
we have performed  numerical
studies of the model (\ref{ind_energy}) at various parameter values. 
{ The computation was conducted as follows: 
First, two phase windings were created around two fixed points
on  a numerical grid.
 Then, the free energy was minimized with 
respect to all degrees of freedom
using a local relaxation method,
constrained so that the vortex cores positions remained fixed.  
The process was then repeated for various separations, 
yielding an intervortex interaction potential.} 

\begin{figure}
\begin{center}
\includegraphics[width=90mm]{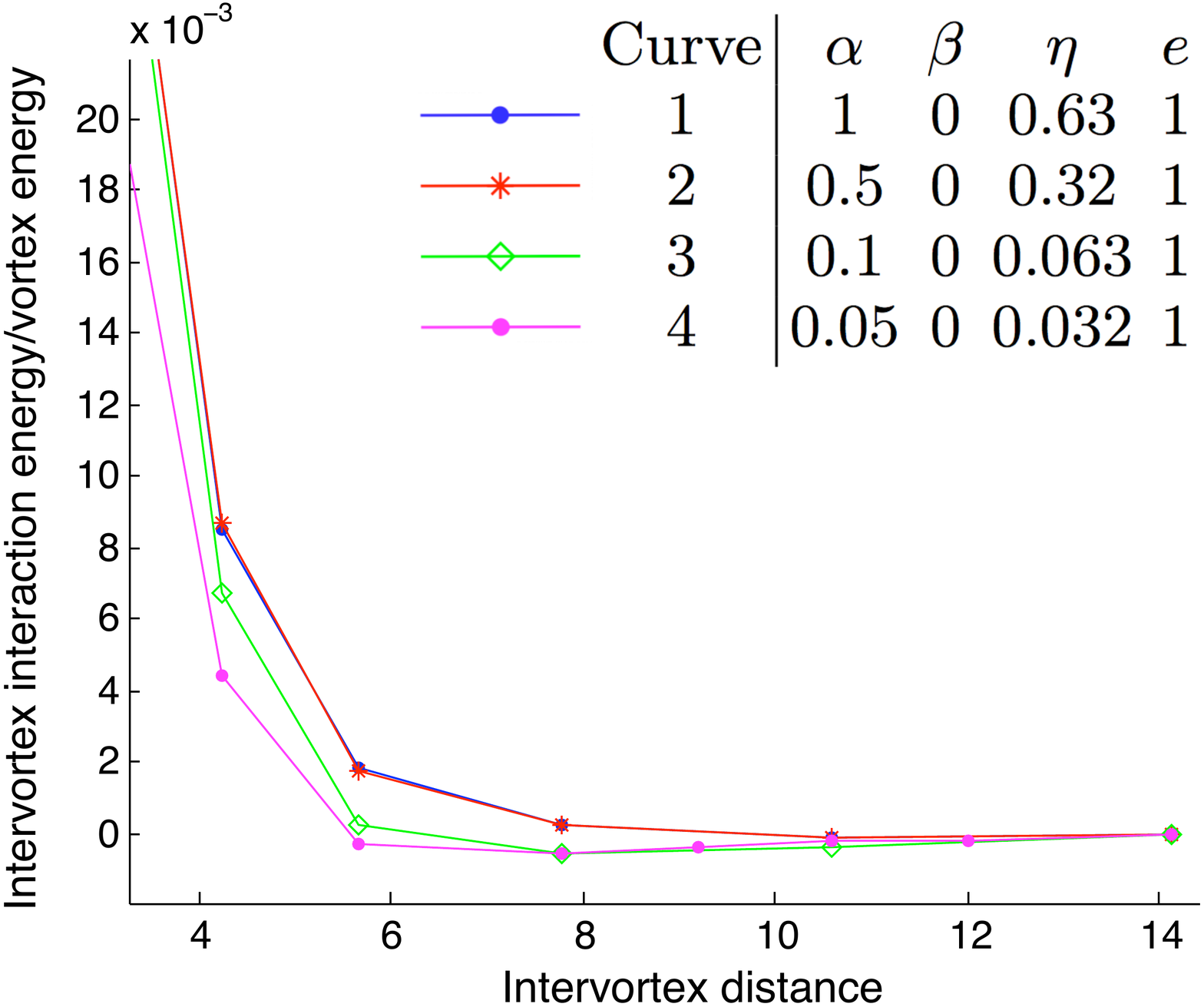}
\end{center}
\caption{(Color online) Interaction energy between two vortices
as a function of vortex separation 
 in units of $10^{-3} E_v$
where $E_v$ is the single vortex energy for a density ratio of $ u_2^2/u_1^2=0.1$.  }
\label{s1}
\end{figure}

\begin{figure}
\begin{center}
\includegraphics[width=90mm]{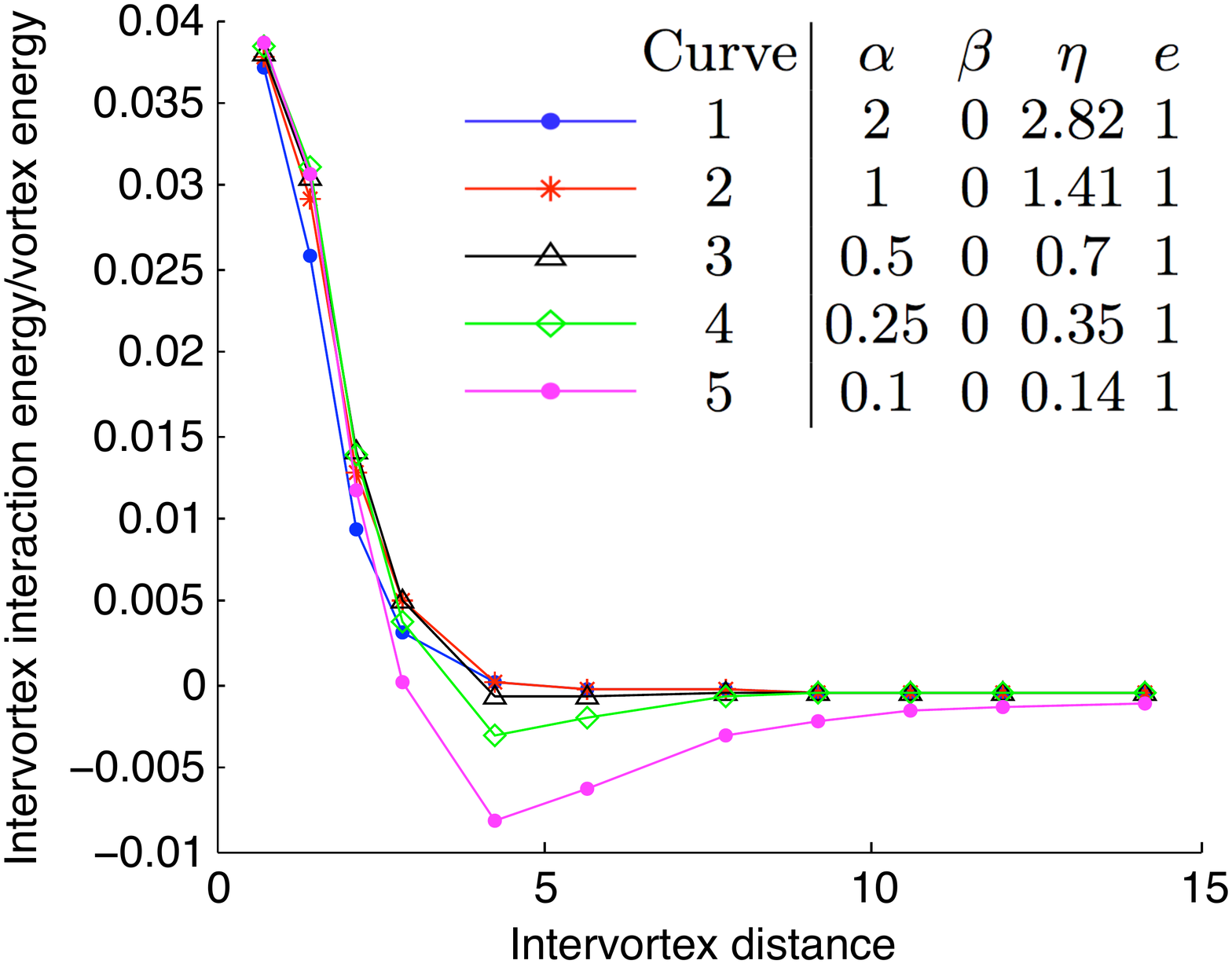}
\end{center}
\caption{(Color online) Interaction energy between two vortices 
as a function of vortex separation for a density ratio of $ u_2^2/u_1^2=0.5$.}
\label{s2}
\end{figure}

\begin{figure}
\begin{center}
\includegraphics[width=90mm]{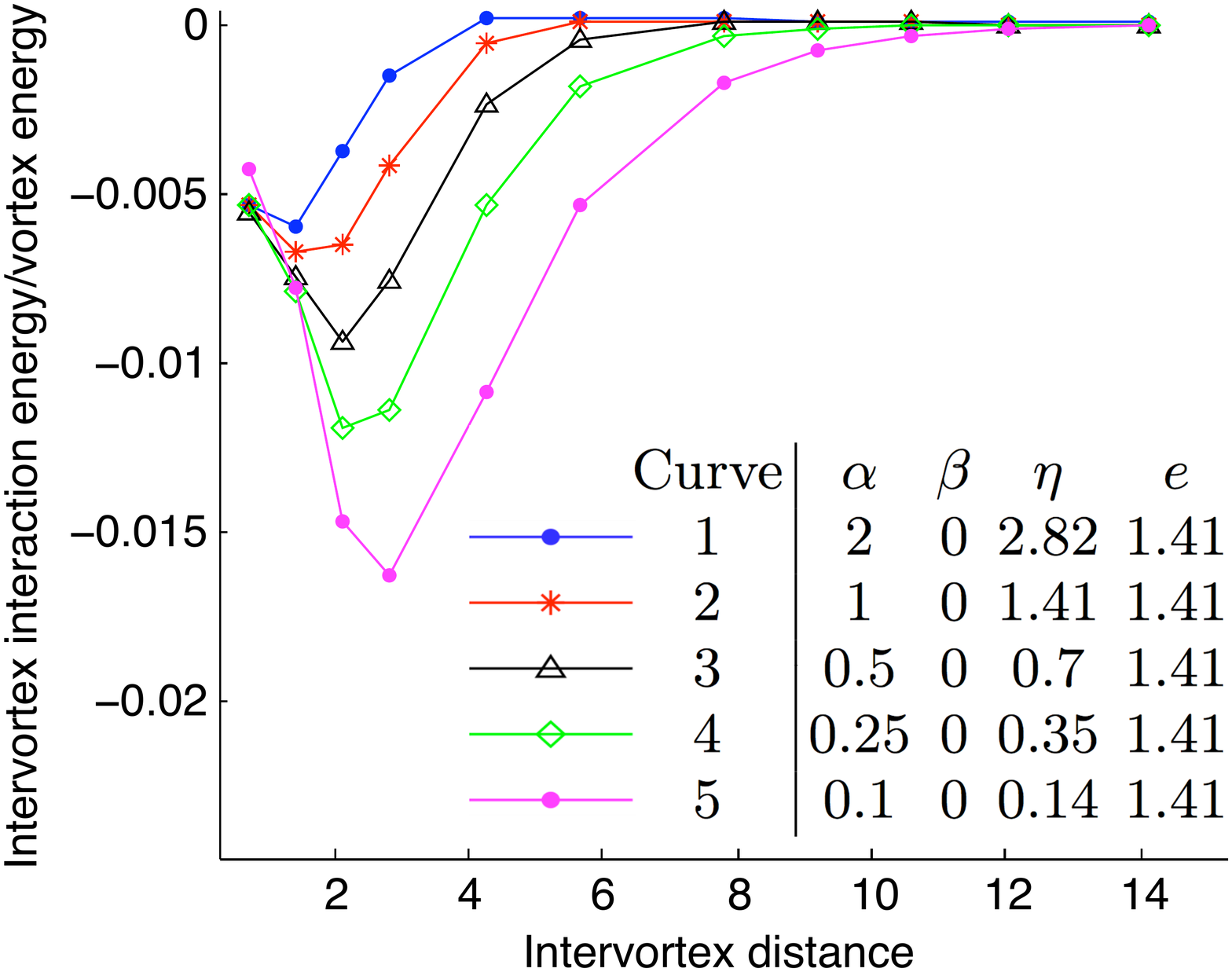}
\end{center}
\caption{(Color online) Interaction energy between two vortices 
as a function of vortex separation for a density ratio of $ u_2^2/u_1^2=0.5$ for $e=1.41$.}
\label{s3}
\end{figure}

First let us consider the regime where the fourth order term in $|\psi_2|$ can be neglected (i.e.\ $\beta=0$).
In this case we conducted computations with the density ratios $|u_2|^2/|u_1|^2$ being $0.1$ and $0.5$.
The results for intervortex interaction energy are presented in Fig. \ref{s1}-\ref{s3}. 
The computed  interaction energy is given in units of $2E_v$ where $E_v$ is the energy of an isolated
single vortex.  
 The length is given in units of  $\sqrt{2}\xi_1$
where $\xi_1$ is a characteristic constant  (the same for all figures)
defined as the  coherence length which can be
associated with this band in the limit of zero coupling to the second band.

In the first case, with density ratio $0.1$  we find that 
in general the { density profiles} of the condensates can be quite different,
even though one  of the bands has proximity-induced superconductivity. This can be ascribed to the fact that the mixing angle $\omega$ is small
(note that $\omega\approx\frac{u_2}{u_1}\frac{2\alpha}{|2\alpha-4|}$) so that the subleading normal mode (of mass $\mu_1>\mu_2$)
dominates $\psi_1$ at intermediate range.
We find that as a consequence of the disparity in the recovery lengths the system 
 crosses over from type-II to type-1.5 behaviour when $\alpha$ and  $\eta$ are sufficiently low
 (Fig. \ref{s1}).  The low density of $\psi_2$
means that the attractive part of the interaction is weak. In the curves 3 and 4, 
we find  a slight long range attraction yielding a minimum energy at around the separation of $r=8$.
%
In the second case (Fig. \ref{s2}), the density ratio is  increased to $0.5$. The vortex-vortex binding energy is now much larger, and the minimum energy occurs at a smaller separation. Long range attraction occur in curves 3-5 with a maximum $\alpha$ of $0.5$, in contrast to   $\alpha \approx 0.1$ in the previous case.

In the third case (Fig. \ref{s3}), the electric charge has been increased by a factor $\sqrt{2}$, 
(which is equivalent to decreasing penetration length) which decreases the magnetic repulsion between vortices. Observe the emergence of a new phenomenon: 
now the energy of an axially symmetric vortex solution 
with two flux quanta is smaller than the energy of two infinitely separated one-quanta vortices. Nonetheless, the axially-symmetric 
two-quantum vortex is not stable since the minimum energy occurs at nonzero vortex separation.

\begin{figure}
\begin{center}
\includegraphics[width=90mm]{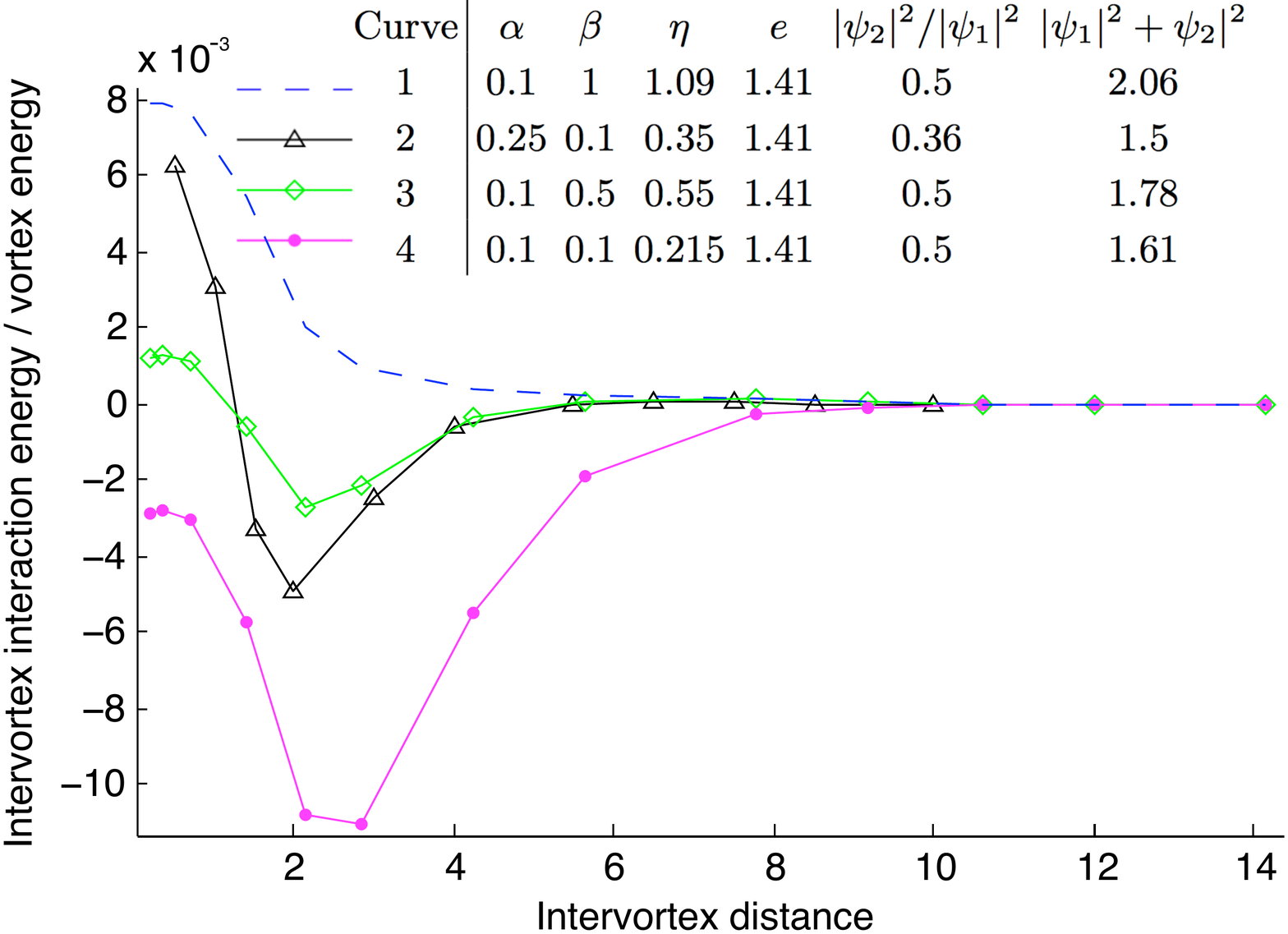}
\end{center}
\caption{(Color online) Intervortex interaction in the presence of fourth-order term for $\psi_2$ in various regimes.}
\label{s5}
\end{figure}

Fig.\ \ref{s5} shows the effect of the addition of a fourth order term with coefficient  $\beta$ in the free energy 
of the proximity-induced component.  It demonstrates the persistence of type-1.5 superconductivity 
when fourth order terms are present for $\psi_2$.
\begin{figure}
\begin{center}
\includegraphics[width=\columnwidth ]{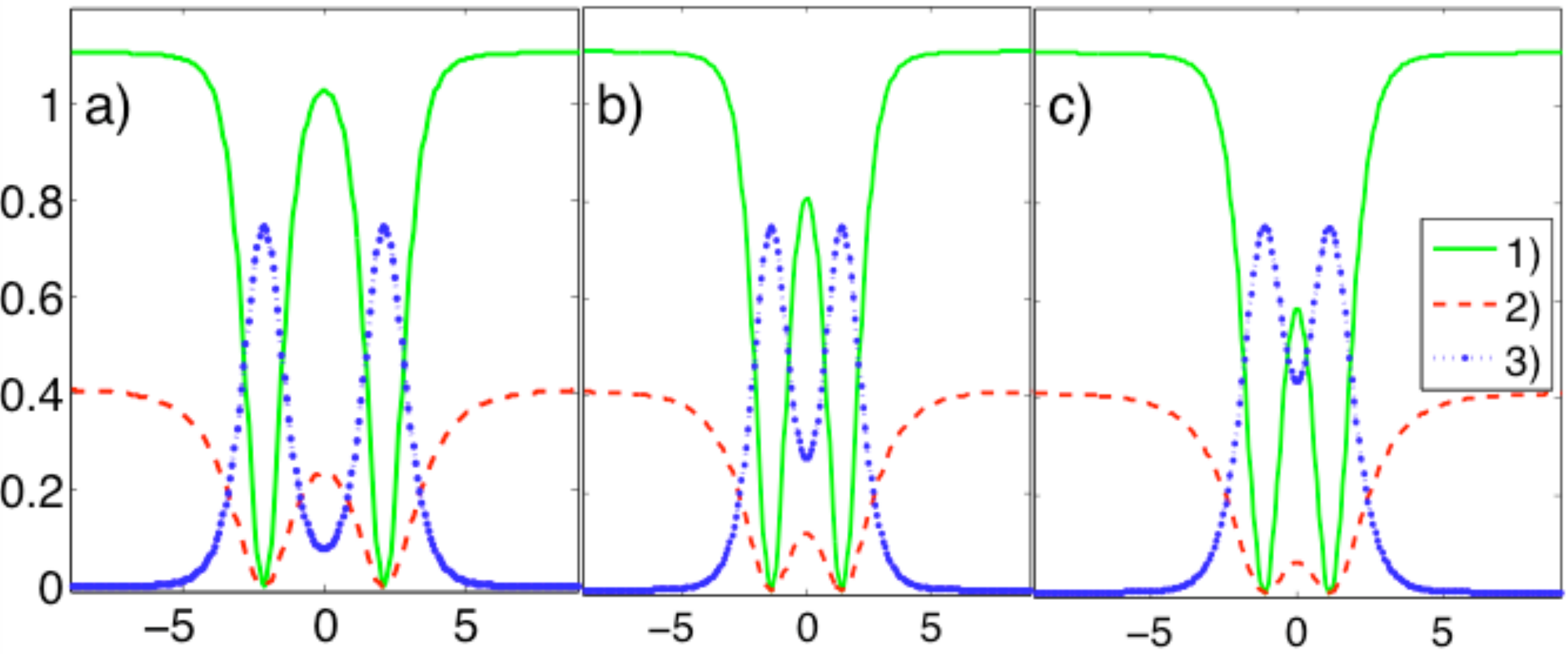}
\end{center}
\caption{(Color online) The behaviour of $|\psi_1|$ (curve 1),
$|\psi_2|$ (curve 2) and magnetic field (curve 3)
for $\alpha=0.25, \eta=0.35, \beta=0.1, e=1.41$.
Separations are (a): 4.24 (corresponding to attractive, $|\psi_2|$-dominated
interaction);  (b): 2.83 (vicinity of the minimum of the interaction potential);
(c):  2.12 (corresponding to  domination of repulsive current-current and 
magnetic interactions). }
\label{cross}
\end{figure}

 To illustrate the actual behaviour of the fields leading to this unusual intervortex
interaction we plot in  fig.\ \ref{cross} cross-sections of the density and magnetic field profile
corresponding to parameter set $2$ in
Fig.\ref{s5}. The figure clearly shows that, in spite of the 
identical long range asymptotics of density behaviour in both bands (as predicted by the linear theory), the rate of
density recovery in both bands at {\it intermediate} scales is actually  different.

{
In conclusion, 
 we considered vortex matter in a  situation which can
take place in two-band systems: only one band is superconducting while
 superfluid density
is induced in another band via an interband proximity effect. This
 situation is in a way antipodal to the previously studied unusual vortex interaction arising in
condensates with independent coherence lengths \cite{bs1},
 e.g.\ as we showed, the  asymptotics of the superfluid
 densities at large distances
 from the core in both  bands are governed by the same exponential law.
 However we find that, in contrast to the conventional
 single-component situation,  the presence of even a tiny interband proximity effect
 can be crucially important. { Namely,  it
gives rise to three fundamental length scales in the problem
and to nontrivial variations of the relative superfluid densities 
in two bands in a vortex
 producing type-1.5 behaviour in a wide range of parameters.
 It should manifest
itself in the magnetic response which involves a phase
separation into vortex and two-component Meissner domains.
The effect may be more common near the temperature
where the weak band crosses over from active to
proximity-induced superconductivity because $\alpha$ should
be small near this temperature.

{ We  thank Alex Gurevich, V. Moshchalkov and M. Wallin for discussions.
The work is supported by the Swedish Research Council,
NSF CAREER Grant No. DMR-0955902, and the U.K.
Engineering and Physical Sciences Research Council.
E. B. was supported by the Knut and Alice Wallenberg
Foundation through the Royal Swedish Academy of
Sciences.}

\end{document}